# Soluble Hydrogen-bonding Interpolymer Complexes in Water: A Small-Angle Neutron Scattering Study

**Maria Sotiropoulou[1], Julian Oberdisse[2] and Georgios Staikos[1]***

[1]*Department of Chemical Engineering, University of Patras, GR-26504 Patras, Greece*
[2]*Laboratoire des Colloïdes, Verres et Nanomatériaux, Université Montpellier II, France and Laboratoire Léon Brillouin CEA/CNRS, CEA Saclay, 91191 Gif sur Yvette, France*

* To whom correspondence should be addressed. E-mail: staikos@chemeng.upatras.gr

ABSTRACT: The hydrogen-bonding interpolymer complexation between poly(acrylic acid) (PAA) and the poly(*N*,*N*-dimethylacrylamide) (PDMAM) side chains of the negatively charged graft copolymer poly(acrylic acid-*co*-2-acrylamido-2-methyl-1-propane sulfonic acid)-*graft*-poly(*N*, *N* dimethylacrylamide) (P(AA-*co*-AMPSA)-*g*-PDMAM), containing 48 wt % of PDMAM, and shortly designated as G48, has been studied by small-angle neutron scattering in aqueous solution. Complexation occurs at low pH (pH < 3.75), resulting in the formation of negatively charged colloidal particles, consisting of PAA/PDMAM hydrogen-bonding interpolymer complexes, whose radius is estimated to be around 165 Å. As these particles involve more than five graft copolymer chains, they act as stickers between the anionic chains of the graft copolymer backbone. This can explain the characteristic thickening observed in past rheological measurements with these mixtures in the semidilute solution, with decreasing pH. We have also examined the influence of pH and PAA molecular weight on the formation of these nanoparticles.

**Introduction**

The formation of hydrogen-bonding interpolymer complexes between weak polyacids, such as poly(acrylic acid) (PAA) or poly(methacrylic acid) (PMAA) and proton acceptor polymers, such as polyethyleneoxide,[1,2,3,4,5,6] polyacrylamides,[7,8,9,10,11] poly(vinylethers)[12,13,14] etc, in aqueous solution has been widely studied in the past four decades. A considerable amount of work on such hydrogen-bonding interpolymer complexes has been presented in two reviews.[15,16] In general, in aqueous solutions of mixtures of such complementary polymers, interpolymer association due to successive hydrogen-bonding between the carboxylic groups of the polyacid and the proton acceptor groups of the polybase leads to the formation of compact interpolymer complexes,[17] soluble only within a narrow pH window. At pH values higher than 4.5-5.5 (where the ionization degree of the weak polyacid is higher than 10–15%), dissociation occurs,[18,19,20,21] while at pH values lower than 3–4 they precipitate as the fraction of the carboxylate anions in the polyacid chain (which is mainly responsible for the solubility of the complex) decreases considerably.[1,2,3,6,11,13,14] The pH dependence of the rate of aggregation, as well as the determination of the size of the aggregate particles has been studied by dynamic light scattering.[22] A small-angle neutron scattering (SANS) study has also directly shown the formation of complexes of a dense structure between PEO and partially neutralized (3 − 9%) PMAA (i.e., in the regime of the formation of soluble hydrogen-bonding interpolymer complexes) in dilute solutions of $D_2O$ and $H_2O$ mixtures.[23]

Extending the solubility of the hydrogen-bonding interpolymer complexes in the low pH region is very important, as this would allow original properties to be observed, thus greatly enlarging the spectrum of the their potential applications. We have recently synthesised an anionically charged graft copolymer, poly(acrylic acid-*co*-2-acrylamido-2-methyl-1-propane sulfonic acid)-*g*-poly (N,N-dimethylacrylamide) (P(AA-*co*-AMPSA)-*g*-PDMAM), by grafting poly (N,N-dimethylacrylamide) (PDMAM) chains onto an acrylic acid-*co*-2-acrylamido-2-methy-1-propane sulfonic acid copolymer (P(AA-*co*-AMPSA)) backbone.[24] PDMAM is a water-soluble polymer with important proton acceptor properties, which forms hydrogen-bonding interpolymer complexes with PAA,[25,26] that precipitate out from water even at pH values as high as 3.75. When these graft copolymers are mixed with PAA in a low pH aqueous solution, hydrogen-bonding interpolymer complexes between the PDMAM side chains and PAA are also formed. Nevertheless, the presence of the negatively charged AMPSA units in the graft copolymer backbone provides these hydrogen-bonding interpolymer complexes with sufficient hydrophilicity, which assures their solubility even at low pH values. At these low pH values, viscosity measurements in dilute solution showed the formation of compact hydrogen-bonding interpolymer complexes. Moreover, rheological measurements in the semi-dilute solution suggest the appearance of a gel-like behaviour; this can be explained by the formation of a network between the negatively charged backbone chains of the graft copolymer, as a result of the hydrogen-bonding interpolymer complexes formed between PAA and the PDMAM side chains of the graft copolymer.[27]

In the present work, SANS experiments have been used to study the microstructure of the above-mentioned electrostatically stabilized colloidal system. An increase in the intensity of the spectrum at low pH values and the appearance of a prominent correlation peak at q ≈ 0.01 Å$^{-1}$ are here attributed to the formation of colloidal nanoparticles; in a semidilute solution, these are functioning as stickers between the extended anionic chains of the graft copolymer backbone, thus leading to the formation of an infinite network. The effect of pH and PAA molecular weight on the structure of these complexes is discussed here in detail.

**Experimental Section**

**Materials.** Four samples of PAA, PAA50, PAA90 (Polysciences), and PAA250, PAA500 (Aldrich), with an average molecular weight of $5.0 \times 10^4$, $9.0 \times 10^4$, $2.5 \times 10^5$ and $4.5 \times 10^5$ g/mol, respectively, were dissolved in a 0.01N HCl solution, dialyzed against water through a cellulose membrane with a molecular weight cutoff equal to 12 kDa (Sigma), and finally obtained by freeze-drying.

The monomers, acrylic acid (AA), 2-acrylamido-2-methyl-1-propane sulfonic acid (AMPSA) (Polysciences), and *N,N*-dimethylacrylamide (DMAM) (Aldrich), were used as received. Ammonium persulphate (APS, Serva), potassium metabisulphite (KBS, Aldrich), 2-aminoethanothiol hydrochloride (AET, Aldrich) and 1-(3-(dimethylamino) propyl)-3-ethyl-carbodiimide hydrochloride (EDC, Aldrich) were used for the synthesis of the graft copolymers.

For the preparation of the buffer solutions, citric acid (CA) and $Na_2HPO_4$ (Merck) were used.

Water was purified by means of a Seralpur Pro 90C apparatus combined with a USF Elga laboratory unit. For the SANS experiments (Aldrich), deuterium oxide was used.

**Polymer synthesis and characterization.** Amine-terminated PDMAM was synthesized by free radical polymerization of DMAM in water at 30 ºC for 6 h using the redox couple APS and AET as initiator and chain transfer agent, respectively. The polymer was purified by dialysis against water through a membrane with a molecular weight cutoff equal to 12.000g/mol (Sigma), and finally obtained by freeze-drying. Its number average molecular weight was determined by end group titration with NaOH after neutralization with an excess of HCl, using a Metrohm automatic titrator (model 751 GPD Titrino) equal to 17.000 g/mol.

A copolymer of AA and AMPSA, P(AA-*co*-AMPSA), was prepared by free radical copolymerization of the two monomers in water, after partial neutralization (90% mole) with NaOH at pH ≈ 6-7, at 30 ºC for 6 h, using the redox couple APS/KBS. The product obtained was then fully neutralized (pH=11) with an excess of NaOH and purified by ultrafiltration with the above Pellicon system and received in its sodium salt form after freeze-drying. Its composition, determined by acid-base titration and elemental analysis, was 18% in AA units. Its apparent weight average molecular weight, $M_w = 2.7 \times 10^5$, was determined by static light scattering in 0.1 M NaCl. As this is a statistical copolymer and the refractive index increment values of its constituents not very different (equal to 0.168 and 0.135, respectively), the error in the molecular weight determination should not be important.

The graft copolymer, P(AA-*co*-AMPSA)-*g*-PDMAM, was synthesized by a coupling reaction between P(AA-*co*-AMPSA) and amine-terminated PDMAM. The two polymers were dissolved in water at a 1:1 weight ratio. Then, an excess of the coupling agent, EDC, was added and the solution was let under stirring for 6 h at room temperature. Addition of EDC was repeated for a second time. The product was purified with a Pellicon system, equipped with a tangential flow filter membrane (Millipore, cut off = 100 kDa), and freeze dried. Its composition in PDMAM side chains was found to be equal to 48 wt% (using elemental analysis), corresponding to about 15 chains per graft macromolecule. A schematic of the graft copolymer is presented in Scheme 1.

**Complex density**. The PAA/PDMAM hydrogen-bonding interpolymer complex was obtained as a precipitate, after mixing equal volumes of a PAA90 7.9 x $10^{-3}$ g/cm$^3$ with a PDMAM 9.9 x $10^{-3}$ g/cm$^3$ aqueous solution, pH = 2.0, and dried in high vacuum for two days. Its density was determined in a methylsalicylate/methylenechloride liquid mixture and was found equal to 1.28 g/cm$^3$.

**Small-angle neutron scattering (SANS).** SANS measurements were carried out at the Laboratoire Léon Brillouin (Saclay, France). The data were collected on beam line PACE at three configurations (6 Å, sample-to-detector distances 1 m and 4.7 m; and 13 Å, 4.7 m), covering a q range from 0.005 to 0.32 Å$^{-1}$ for the semidilute solutions, while the (20 Å, 4.7 m) configuration, covering the very low q range (of 0.0034 to 0.022 Å), was also used for the dilute solutions. 2 and 5 mm light path quartz cells for the semidilute and dilute solutions, correspondingly, were used. Empty cell scattering was subtracted and the detector was calibrated with 1 mm H$_2$O scattering. All measurements were carried out at room temperature. Data were converted to absolute intensity through a direct beam measurement, and the incoherent background was determined with H$_2$O/D$_2$O mixtures.

**Preparation of the polymer mixture solutions.** The solutions used were either semidilute or dilute solutions of mixtures of the graft copolymer G48 with the polyacids PAA50, PAA90, PAA250 and PAA450 in D$_2$O, at different pH values, adjusted with citric acid-phosphate buffers. For the relatively low ionic strength of the solutions used, the crossover concentration between the dilute and the semidilute regime has been viscometrically estimated to be in the region (4 – 6) x $10^{-3}$ g/cm$^3$ of G48. The concentration of G48 in each semidilute solution was constant, equal to 1.40 x $10^{-2}$ g/cm$^3$, while the quantity of the polyacid (i.e., of PAA) was varying, depending on the polyacid/polybase unit mol ratio measured (polybase is the PDMAM grafted on the P(AA-*co*-AMPSA) backbone of G48). The dilute solutions used were prepared through a five times dilution of the semidilute solutions with D$_2$O buffer solutions. The pH of the solutions was checked with a Metrhom pHmeter equipped with a combined pH glass electrode. The solutions after their preparation were let under stirring for 24 hours at room temperature.

**Results and Discussion**

**Semidilute solutions.** Figure 1 shows the variation of the SANS intensity, I, versus the scattering wave vector, q, for pure G48 and PAA90, and for G48/PAA90 mixtures, at four different unit mole polyacid / polybase ([PAA90] / [PDMAM]) ratios, r = 0.25, 0.5, 1, 1.5, in $D_2O$, at pH=2. Even if G48 is a polyelectrolyte, it does not exhibit a structure factor peak in the region of q values between $10^{-2}$ and $10^{-1} Å^{-1}$, probably due to the relatively high ionic strength that results from the presence of citric acid to the buffer pH 2.[28, 29] PAA90 (a weak polyelectrolyte, practically neutral at this pH), exhibits also a typical (for polymers) neutron scattering behavior. Nevertheless, as some PAA90 is added in the G48 solution (for instance at r = 0.25), a noteworthy increase in the scattering intensity is observed in the low q region, characteristic of a system structured at the scale 1/q, i.e. approximately 100 Å. The only way to explain the peak in this rather low q range is to assume the formation of bigger objects, formed by complexation. Anyway, in this low pH region, insoluble hydrogen–bonding interpolymer complexes are known to be formed between PDMAM and PAA. The scattering intensity, I, increases further, as r increases up to r = 1, while remains constant by further adding PAA90. The behavior observed is explained by the hydrogen-bonding complexation of PAA90 with the PDMAM side chains grafted onto the anionic P(AA-*co*-AMPSA) backbone of the graft copolymer, G48. As a result, compact colloidal complex particles, stabilized by the anionic chains of the graft copolymer backbone, are formed, schematically presented in Scheme 2, that are evidenced by this high scattering.

The low q intensity corresponds to a first approximation to the Guinier regime of the scattering of individual non-interacting, dense objects;[30] their radius leads to a characteristic decrease in I, whose magnitude is related to their mass. If the objects are spheres of radius R:

$$I = I_o \exp(-R^2 q^2/5) \qquad (1a)$$

with

$$I_o = \varphi \, \Delta\rho^2 \, V_o \qquad (1b)$$

where Vo denotes the volume of an individual object, φ the volume fraction of the objects and Δρ the scattering contrast. In the present case, these objects appear to interact presumably due to electrostatic charges, which introduce some structure, namely an intensity peak around q ≈ 0.01 Å$^{-1}$. At high q, the same scattering intensity is found as for pure G48 solution.

As indicated above, even after a partial complexation of G48 with PAA90, i.e., in the presence only of some, not too strong yet, repulsive structure factor, the scattering intensity exhibits a maximum, related to the mass of the complexes. In Figure 2 the maximum of the scattering intensity, $I_{max}$, is plotted as a function of the unit mole ratio, r, for the G48/PAA90 polymer mixtures in the semidilute regime, at pH 2. $I_{max}$ is observed to increase with increasing r, until r = 1.1, which should correspond to the stoichiometry of the hydrogen-bonding interpolymer complex that is formed between PAA90 and PDMAM; then, a plateau is practically attained.

*Solution structure: estimation based on a simple cubic lattice model.* The information provided by the structure factor is exploited in Figure 3, where the variation of I with q is presented for the G48/PAA90 mixture with r = 1.1, in a D$_2$O semidilute solution, at pH 2. A structure peak at $q_0$ = 0.0095 Å$^{-1}$ shows up corresponding to the most probable distance of the particles, consisted of the compact hydrogen-bonding interpolymer complexes formed between the PAA90 chains and the PDMAM side chains of the G48 graft copolymer. This information can be used to estimate the "dry" radius, $R_{dry}$, of the particles, by using a simple cubic lattice model (CLM) based on the mass conservation of the polymer, with the distance between the particles given through D = 2π/$q_0$. $R_{dry}$ corresponds to the radius of the complex particles with all solvent molecules being expelled. Since the volume of each particle can be estimated through V = φD$^3$, where φ is the volume fraction of the complex particles, $R_{dry}$ can be calculated by the equation

$$R_{dry} = \sqrt[3]{\frac{6\pi^2 \varphi}{q_0^3}} \qquad (2)$$

For φ equal to 9.5x10$^{-3}$ (calculated by taking into account the concentration of the complex particles formed by the hydrogen-bonding interaction of the PDMAM side chains of G48 with the PAA90 chains and the mass density of the complex), $R_{dry}$ comes out to be equal to 87 Å. From $R_{dry}$ we can also obtain the molecular mass, $M_c$, of the dry complex particle through

$$M_c = (4/3)\pi R_{dry}^3 d_c N_A \qquad (3)$$

where $N_A$ is Avogadro's number. Eq. (3) gives $M_c = 2.1 \times 10^6$ Da. Based on the complex composition, this implies that each particle should contain about 70 PDMAM side chains, corresponding to at least 5 graft copolymer chains. That is, for every complex particle to be formed at least five graft copolymer chains should be involved. This inevitably leads to the formation of a network and explains the noteworthy thickening behavior that has been observed with such mixtures in semidilute aqueous solution at low pH values.[24, 27]

The intensity at q → 0 could be estimated from Eq. (1b), where $V_o = 4\pi R_{dry}^3/3$. The scattering contrast, Δρ, was equal to $5.0 \times 10^{10}$ cm$^{-2}$, as $\Delta\rho = \rho_s - \rho_c$, where $\rho_s$ is the neutron scattering length density of the solvent (D$_2$O), $6.4 \times 10^{10}$ cm$^{-2}$, and $\rho_c$ this one of the dry complex (PAA/PDMAM, with a 1.1/1 unit mol ratio), $1.4 \times 10^{10}$ cm$^{-1}$. Then, Eq. (1b) with the CLM gives the theoretical value $I_{o,th} = 64$ cm$^{-1}$, not very far from the value of the 95 cm$^{-1}$, which, as we will see shortly, is the value of the real prefactor of the form factor.

*Solution structure: estimation using the rescaling mean spherical approximation.* We will now try to reproduce the main features of the observed intensity pattern with a simple structural model, in order to check if our data interpretation is indeed compatible with the experimental observations. The model is based on a simple form factor for the complexes and a repulsive interaction between charged particles. The form factor has been estimated by fitting the Guinier-law at intermediate angles, i.e., at $q > q_0$. Its zero angle intensity is 95 cm$^{-1}$, and the corresponding radius is 165 Å. An effective Yukawa potential can be used for the description of the repulsive structure factor. The corresponding quasi-analytical structure factor has been calculated by Hayter and Penfold (mean spherical approximation),[31] with the rescaling proposed by Hansen and Hayter (RMSA).[32] The parameters are the effective charge on the complex, its dry radius, the Debye-Hückel length (characterizing the length scale of the interactions in the solution), and the complex volume fraction. The features which need to be reproduced are the osmotic compressibility (given by the low-angle limit of the structure factor), and of course the peak position and height, after they have all been multiplied with the above-mentioned structure factor. In order to obtain the experimental peak position which is determined by the degree of repulsion between complexes and the mass conservation, the dry complex radius needs to be set to 100 Å. According to Eq. 1(b), this corresponds to $I_o = 99$ cm$^{-1}$, and is thus consistent with the form factor fit (95 cm$^{-1}$). We have then fixed the Debye-Hückel length to 45 Å, and varied the charge in order to achieve a satisfying fit of the total intensity.

A good fit with a value for the charge equal to 70 is shown in Figure 3, where in the high q region we have also included a power law scaling of I with q of the form $I \sim q^{-d}$ to account for the polymer scattering, with a value for the exponent d equal to 1.6 (i.e., close to 5/3 characteristic of a chain with excluded volume).[33] The good overall agreement indicates that the description is not too far from reality.

It is interesting to note that the wet radius given by the form factor (which corresponds to the real, hydrated size of the complexes) is considerably larger than the dry one. This means that much of the volume of each complex particle is occupied by water, probably explaining why the polymer chain scattering (high q region) is almost unchanged upon complex formation. The number of charges per complex particle is difficult to determine. This charge is provided by the AMPS units of the anionic graft copolymer backbone, which is not participating in the formation of the dense hydrogen-bonding complex particles, but plays an important role in the stabilization of the particles. Since about 70 PDMAM chains are involved in the formation of each particle (and since they are grafted onto the anionic backbone of AMPS units), at least two of them (the first neighbors) should be considered as a part of the particle surface. It means that a number of 140 could be realistic for the particle charge. Nevertheless, as this charge is along a polyelectrolyte chain, ion condensation should be taken into account resulting in a considerably lower number. Due to these reasons, the value estimated by the RMSA structure factor seems to be realistic. Note that the scope of the RMSA procedure was to find a satisfying description of the structure factor in order to have access to the shape of the complexes. The solution found is not unique, and a smaller Debye length (e.g., equal to 40 Å, i.e., closer to the 30 Å estimated from the ionic strength) with a higher charge (e.g., 90) reproduces the scattering just as well. The order of magnitude, however, remains unchanged, and stays compatible with the values of the parameters deduced from the physical chemistry.

*pH dependence.* In Figure 4(a), we present the intensity I vs. q for the G48/PAA90 mixture with r = 1.10, at five different pH values. We observe that at high enough pH values (e.g., pH = 4.2) where the hydrogen-bonding complexation between G48 and PAA90 is prevented due to the negative charge of the partially neutralized PAA90 chain, the scattering curve is representative of a macromolecular chain in solution, with maximum scattering intensity lower than 1 cm$^{-1}$ and a q exponent close to 1, corresponding to a rigid rod, probably due to the expanded form of PAA at this pH.

As pH decreases (e.g., down to pH = 3.5), the intensity at low q values is observed to increase by almost by one order of magnitude (down to around 9 cm$^{-1}$), indicating a chain organization in the system. In the high q region, a higher q exponent is obtained (d = 1.7), corresponding to a polymer coil in a good solvent (chain with excluded volume). At this pH (which is lower than the critical pH value, $pH_c$ = 3.75), a hydrogen-bonding interpolymer complex between PAA90 and the PDMAM side chains of the G48 graft copolymer is formed, explaining the observed behavior. By further decreasing the pH value, the intensity at low q values increases further reaching a peak of 62 cm$^{-1}$ at q = 0.0095 Å$^{-1}$ for pH = 2. It is also observed that the values of the scattering intensity for pH = 2.5 and pH = 3.0 are not much different, which indicates that at pH = 3.0 the complex formed has almost assumed its final form so that by further decreasing the pH value no considerable changes take place.

In Figure 4(b) the maximum of intensity, $I_{max}$, is plotted as a function of pH. A sigmoidal curve is found to fit the experimental data quite well, showing that at around pH equal to 3.5, a transition occurs. Its origin should be related to the formation of the interpolymer complex due to hydrogen-bonding, and the structural organization of the system. In this pH region, the scattered intensity increases by about two orders of magnitude as the value of the pH decreases from about 4 to about 3.

*Influence of the molecular weight of PAA.* Figure 5 shows the scattering curves obtained with mixtures of G48 and four PAA samples of different molecular weights, ranging from 50 000 Da (PAA50) to 450 000 Da (PAA450), with r = 1.10, at pH = 2.0. We observe that, the four curves for all the mixtures with different molecular weights of the PAA samples practically coincide, indicating that the size of the particles formed is not influenced by the molecular weight of the weak polyacid. Nevertheless, interconnectivity should change, since by increasing the PAA molecular weight, the viscosity and the reptation time also increase.

**Dilute solutions.** The scattering curve of the G48/PAA90 mixture in a $D_2O$ dilute solution (obtained from the corresponding semidilute solution with r = 1.10, at pH = 2.0, after a five times dilution), is shown in Figure 6(a) over a broad range of q values. Also shown in the Figure is the curve corresponding to the semidilute mixture (initially shown in Figure 3), after dividing the scattered intensity by the dilution ratio, i.e., by 5. We see that the two curves practically coincide, except from some minor deviations at large q values due to the difficult background subtraction, and the presence of the low q structure peak at the higher concentration. This means that the size of the formed particles is not influenced by the concentration of the solution, or, equivalently, the structure of the complex particles does not change with concentration.

Moreover, a model RMSA-calculation for the dilute solution data with the same parameters as for the semidilute solution (besides the concentration) has been performed. The result is expressed by the continuous line in Figure 6(b), and shows a satisfying agreement with the experimental data in the (relevant) low q region. The only notable difference is in the $I_o$ value, which is now 16, i.e., not far from 95/5 = 19, that is as expected after the five times dilution. Naturally, minor changes in shape cannot be excluded, but this calculation demonstrates that our approach is self-consistent.

Figure 7 shows I vs. q curves in the very low q region, for mixtures of the G48 graft copolymer with four PAA samples of different molecular weight, PAA50, PAA90, PAA250 and PAA450, for r = 1.10, in a dilute $D_2O$ solution, at pH = 2.0. We observe that all the systems present almost the same scattering behavior, indicating that differences in the PAA molecular weight do not considerably influence the size of the particles formed. This behavior coincides with the behavior exhibited by the semidilute solutions presented in Figure 5.

**Conclusions**

We have studied the hydrogen-bonding interpolymer complexation between PAA and PDMAM grafted onto a negatively charged backbone (P(AA-*co*-AMPSA)) by SANS measurements. Our results revealed that for pH values lower than 3.75, a structured system is formed consisting of colloidal particles; the latter are comprised by a compact core of PAA/PDMAM hydrogen-bonding interpolymer complexes and surrounded by anionic hydrophilic P(AA-*co*-AMPSA) chains. These complexes, playing the role of stickers, lead to the formation of a physical network in semidilute solution, explaining their previously observed gel-like behavior.[24, 27] In addition, it was shown that their radius (which for pH=2 was estimated to be equal to 165 Å) is not influenced by the molecular weight of PAA; their structure was also found not to change with concentration.

**Acknowledgment.** We thank the European Social Fund (ESF), the Operational Program for Educational and Vocational Training II (EPEAEK II) and particularly the Program IRAKLEITOS, for funding the present work. This research project has also been supported by the European Commission under the 6th Framework Programme through the Key Action: Strengthening the European Research Area, Research Infrastructures. Contract n°: HII3-CT-2003-505925.

**Scheme 1.** A schematic depiction of the graft copolymer P(AA-*co*-AMPSA)-*g*-PDMAM (G48).

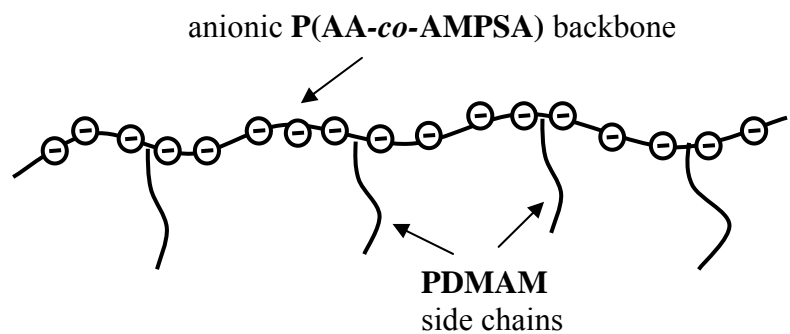

anionic **P(AA-*co*-AMPSA)** backbone

**PDMAM** side chains

**Scheme 2.** Negatively charged colloidal particles formed through hydrogen-bonding interpolymer complexation of PAA with the PDMAM side chains of the graft copolymer P(AA-*co*-AMPSA)-*g*-PDMAM (G48), at low pH.

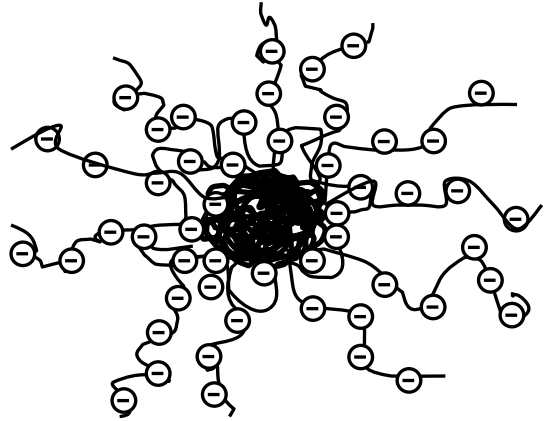

# Figure Captions

**Figure 1.** SANS intensity, I, vs. wave vector q, for pure PAA90, (○); G48 (●) and G48/PAA90 polymer mixtures of different polyacid/polybase unit mole ratios, r = [PAA]/[PDMAM], 0.25, (■); 0.5, (Δ); 1, (▲); 1.50, (□) in semidilute solutions in $D_2O$, at pH=2. The concentration of G48 is $1.40 \times 10^{-2}$ g/cm$^3$, both in the pure G48 solution and in its mixtures with PAA90. The concentration of the pure PAA90 solution is $1.00 \times 10^{-2}$ g/cm$^3$.

**Figure 2.** Variation of the maximum in the scattering intensity, $I_{max}$, for the G48/PAA90 polymer mixtures in semidilute solution in $D_2O$, at pH = 2, vs. their unit mole ratio r = [PAA90]/[PDMAM].

**Figure 3.** Variation of I vs. q for the G48/PAA90 polymer mixture in a semidilute solution in $D_2O$, for r = 1.1 and pH=2.0: experimental points and an "RMSA + high q tail" fit.

**Figure 4. (a)** I vs. q for the G48/PAA90 mixture in semidilute solution in $D_2O$, for r = 1.10, at five different pH values: pH = 4.2 (●); pH = 3.5 (■); pH = 3.0 (○); pH = 2.5 (▲); pH = 2.0 (□);. **(b)** $I_{max}$ vs. pH for the same mixture.

**Figure 5.** I vs. q, for mixtures of the graft copolymer G48 with four different molecular weight PAA samples, PAA50, (■); PAA90, (●); PAA250, (▲); and PAA450, (□) in semidilute solution in $D_2O$, for r = 1.10 and pH = 2.0

**Figure 6. (a).** I vs q for the G48/PAA90 mixture in a dilute solution in $D_2O$, for r = 1.1 and pH = 2.0: (●), and superposition of the corresponding semidilute solution curve (originally shown in Figure 3) after dividing the scattering intensity I values by 5, i.e., the dilution ratio: (○).
**(b).** I vs q for the dilute solution of the G48/PAA90 mixture of Figure 6(a) in the low q region. Experimental points and an "RMSA" fit.

**Figure 7.** I vs. q in the low q region for the G48/PAA50 mixtures, (■); G48/PAA90, (●); G48/PAA250, (▲); G48/PAA450, (□); with r = 1.10 in dilute solution in $D_2O$ and for pH = 2.0.

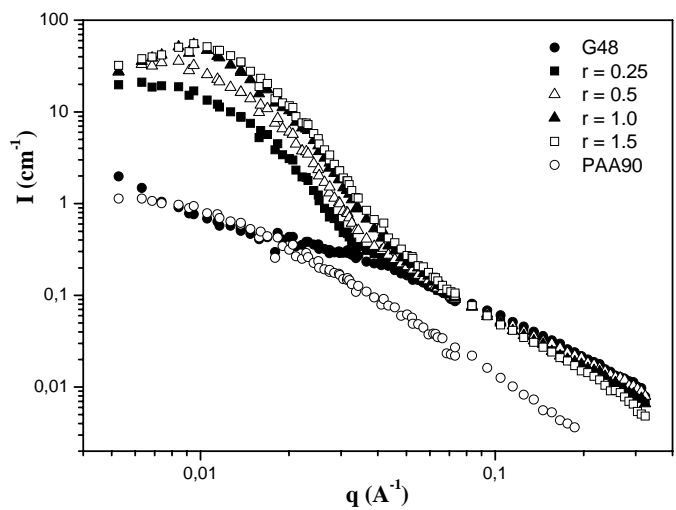

**Figure 1**

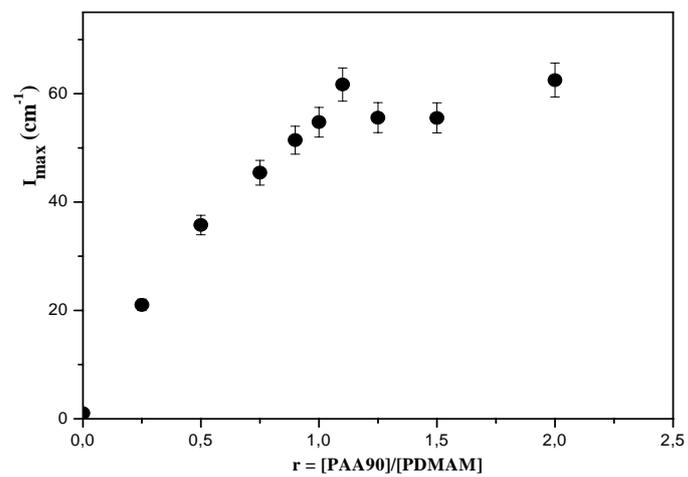

**Figure 2**

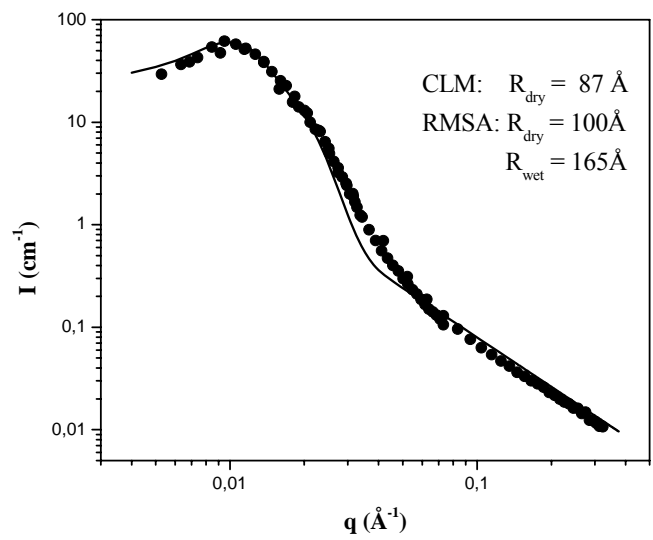

**Figure 3**

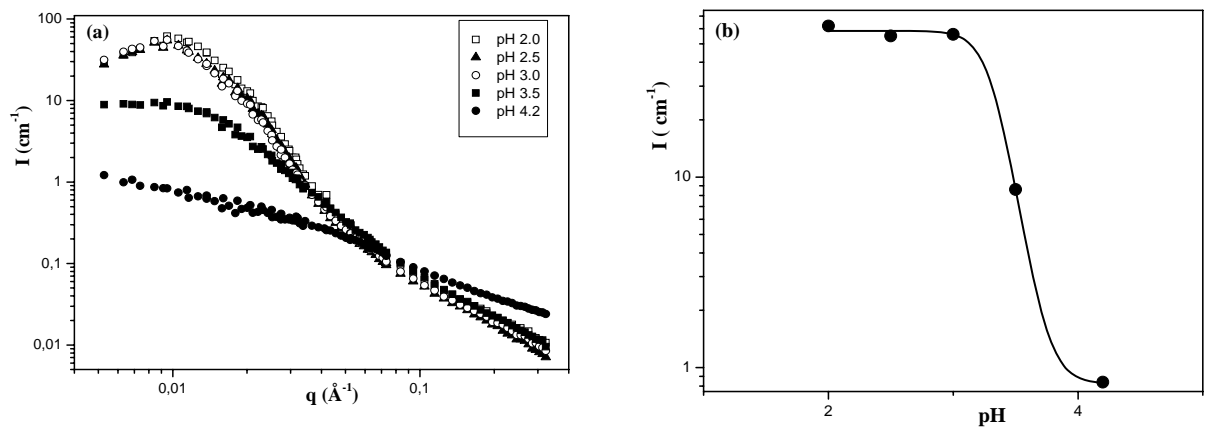

**Figure 4**

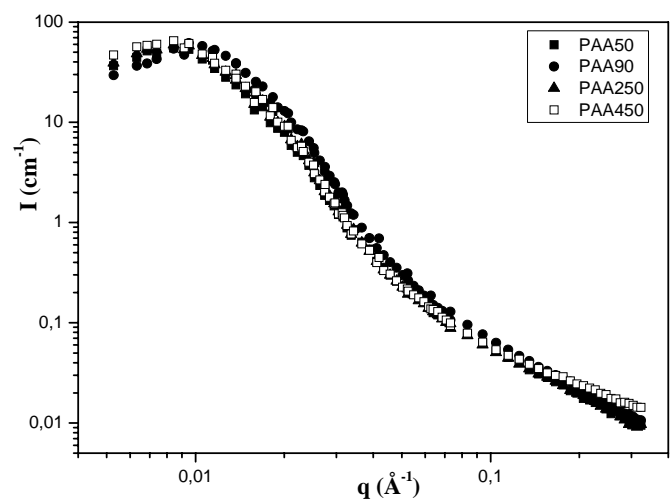

**Figure 5**

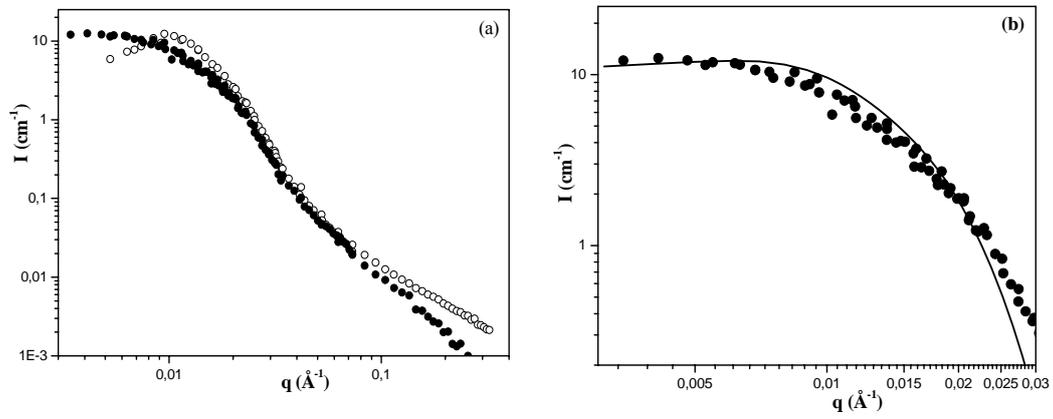

**Figure 6**

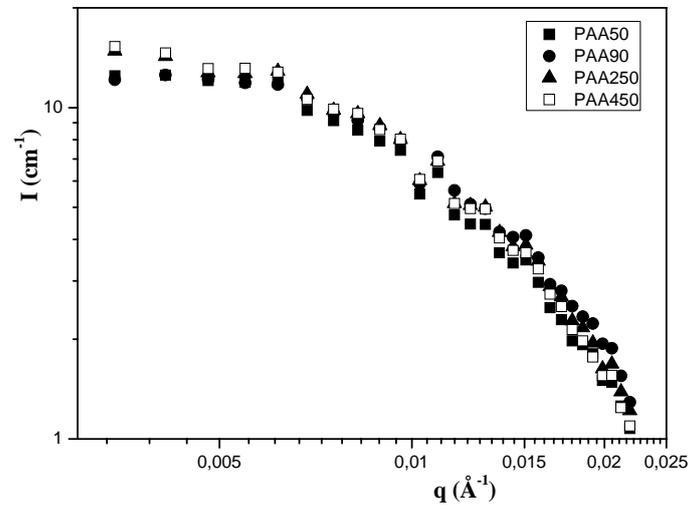

**Figure 7**

**Soluble Hydrogen-bonding Interpolymer Complexes in Water: A Small-Angle Neutron Scattering Study**

Maria Sotiropoulou, Julian Oberdisse and Georgios Staikos[*]

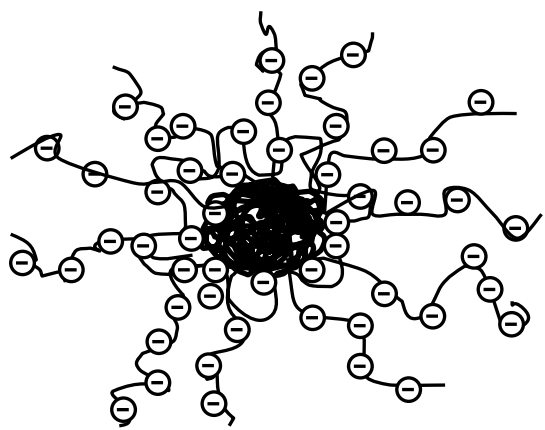
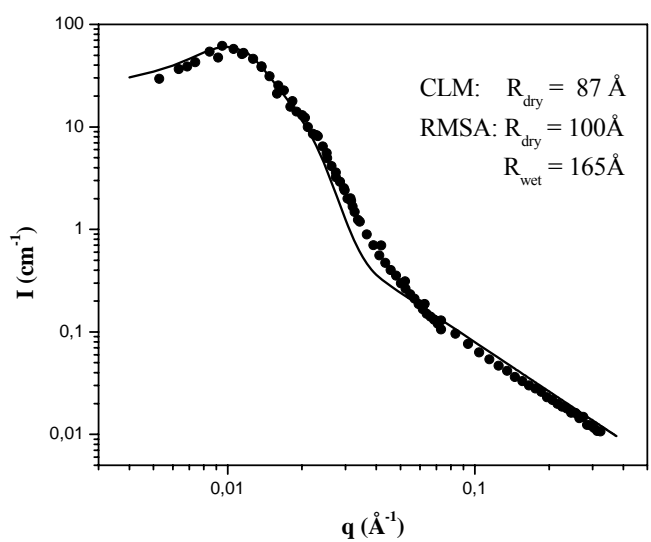

**References**

(1) Bailey, F.E.; Lundberg, JR., R.D.; Callard, R.W. *J. Polym. Sci. Part A* **1964**, *2*, 845.

(2) Ikawa, T.; Abe, K.; Honda, K.; Tsuchida, E. *J. Polym. Sci.Polym. Chem. Ed.* **1975**, *13*, 1505.

(3) Ohno, H.; Matsuda, H.; Tsuchida, E. *Makromol. Chem.* **1981**, *182*, 2267.

(4) Iliopoulos, I.; Audebert, R. *Polym. Bull.* **1985**, *13*, 171.

(5) Hemker, D.J.; Garza, V.; Frank, C.W. *Macromolecules* **1990**, *23*, 4411.

(6) Khutoryanskiy, V. V.; Dubolazov, A. V.; Nurkeeva, Z. S.; Mun, G. A. *Langmuir* **2004**, *20*, 3785.

(7) Klenina, O.V.; Fain, E.G. *Polym. Sci. U.S.S.R.* **1981**, *23*, 1439.

(8) Eustace, D.J.; Siano, D.B.; Drake, E.N. *J. Appl. Polym. Sci.* **1988**, *35*, 707.

(9) Staikos, G.; Karayanni, K.; Mylonas, Y. *Macromol. Chem. Phys.* **1997**, *198*, 2905.

(10) Aoki, T.; Kawashima, M.; Katono, H.; Sanui, K.; Ogata, N.; Okano, T.; Sakurai, Y. *Macromolecules* **1994**, *27*, 947.

(11) Mun, G. A.; Nurkeeva, Z. S.; Khutoryanskiy, V. V.; Sarybayeva, G. S.; Dubolazov, A. V. *Eur. Polym. J.* **2003**, *39*, 1687.

(12) Karayanni, K.; Staikos, G.; *Eur. Polym. J.* **2000**, *36*, 2645.

(13) Mun, G. A.; Nurkeeva, Z. S.; Khutoryanskiy, V. V.; Sergaziyev, A. D. *Colloid Polym. Sci.* **2002**, *280*, 282.

(14) Nurkeeva, Z. S.; Khutoryanskiy, V. V.; Mun, G. A.; Bitekenova, A. B. *Polym. Sci.Ser. B* **2003**, *45*, 365.

(15) Bekturov, E. A.; Bimendina L.A. *Adv. Polym. Sci.* **1980**, *43*, 100.

(16) Tsuchida, E.; Abe, K. *Adv. Polym. Sci.* **1982**, *45*, 1.

(17) Sivadasan, K.; Somasundaran, P.; Turo; N.J. *Coll. Polym. Sci.* **1991**, *269*, 131.

(18) Iliopoulos, I.; Audebert, R. *Eur. Polym. J.* **1988**, *24*, 171.

(19) Oyama, H.T.; Hemker, D.J.; Frank, C.W. *Macromolecules* **1989**, *22*, 1255.

(20) Iliopoulos, I.; Audebert, R. *Macromolecules* **1991**, *24*, 2566.

(21) Bokias, G.; Staikos, G.; Iliopoulos, I.; Audebert, R. *Macromolecules* **1994**, *27*, 427.

(22) Usaitis, A.; Maunu, S. L.; Tenhu, H. *Eur. Polym. J.* **1997**, *33*, 219.

(23) Zeghal, M.; Auvray, L. *Europhys. Lett.* **1999**, *45*, 482.

(24) Sotiropoulou, M.; Bokias, G. ; Staikos, G. *Macromolecules*, **2003,** *36*,1349.

(25) Wang, Y.; Morawetz, H. *Macromolecules* **1989**, *22*, 164.

(26) Shibanuma, T.; Aoki, T.; Sanui, K.; Ogata, N.; Kikuchi, A.; Sakurai, Y.; Okano, T. *Macromolecules* **2000**, *33*, 444.

(27) Bossard, F.; Sotiropoulou, M.; Staikos, G. *J. Rheol.* **2004**, *48*, 927.

(28) Koyama, R. *Macromolecules* **1984**, *17*, 1954.

(29) Morfin, I.; Reed, W.F.; Rinaudo, M.; Borsali, R. *J. Fhys. II France* **1994**, *4*, 1001.

(30) Lindner, P.; Zemb, Th., Eds.; *Neutrons, X-rays and Light: Scattering Methods Applied to Soft Matter*; North-Holland, Delta Series, Elsevier: Amsterdam, 2002.

(31) Hayter, J.B.; Penfold, J. *Mol. Phys.* **1981**, *42*, 109.

(32) Hansen, J.P.; Hayter, J.B. *Mol. Phys.* **1982**, *46*, 651.

(33) Higgins, J. S.; Benoît, H. C. In *Polymers and Neutron Scattering*; Oxford Science Publications, Clarenton Press: Oxford, 1994.